Published manuscript in Springer Nature GPS Solutions.

Content courtesy of Springer Nature, terms of usage apply. Rights reserved. This work is published in Springer Nature GPS Solutions under Creative Commons Attribution License CC-BY

This publication is **Open Access (OA)** and free for downloading at https://link.springer.com/article/10.1007/s10291-023-01447-z. Please cite this article as:

Liu, S., Wang, K. & Abel, D. Robust state and protection-level estimation within tightly coupled GNSS/INS navigation system. GPS Solut 27, 111 (2023). https://doi.org/10.1007/s10291-023-01447-z.

**Robust State and Protection-Level Estimation within Tightly Coupled GNSS/INS Navigation System**

**Shuchen Liu*[1], Kaizheng Wang[1] and Dirk Abel[1]**

[1]*Institute of Automatic Control, RWTH Aachen University, Germany*

**Corresponding author* (E-mail: s.liu@irt.rwth-aachen.de)

## Abstract

In autonomous applications for mobility and transport, a high-rate and highly accurate vehicle-state estimation is achieved by fusing measurements of global navigation satellite systems (GNSS) and inertial sensors. The state estimation and its protection-level generation often suffer from satellite-signal disturbances in urban environments and subsequent poor parametrization of the satellite observables. Thus, we propose an innovative scheme involving an extended $H_\infty$ filter (EHF) for robust state estimation and zonotope for the protection-level generation. This scheme is shown as part of a tightly coupled navigation system based on an inertial navigation system and aided by the GPS/Galileo dual-constellation satellite navigation system. Specifically, GNSS pseudorange and deltarange observables are utilized. The experimental results of post-processing a real-world dataset show significant advantages of EHF against a conventional extended Kalman filter regarding the navigation accuracy and robustness under various GNSS-measurement parametrizations and environmental circumstances. The zonotope-based protection-level calculation is proven valid, computationally affordable, and feasible for real-time implementations.

## Introduction

Recently, considerable efforts have been made in autonomous driving, as it offers great potential in terms of safety and efficiency. For autonomous applications, navigation systems are indispensable, which shall robustly and accurately estimate the vehicle states, namely position, velocity, and orientation. Further, integrity is also considered an essential requirement for navigation systems in safety-critical applications (Bhatti and Ochieng 2007). Often, a high-rate and highly accurate vehicle-state estimation is achieved by fusing measurements of global navigation satellite systems (GNSS) and inertial navigation systems (INS) using Kalman filtering. In tightly coupled navigation systems, Kalman filter solutions cannot guarantee a reliable vehicle-state estimation under inexact noise statistics. In this case, an extended $H_\infty$ filter (EHF) can be used alternatively, outperforming the Kalman filtering regarding system uncertainties (Simon 2006).

Moreover, facing various software and hardware failures in real-world operations (Bhatti and Ochieng 2007), navigation filters need to be extended with an integrity monitoring system. Besides detecting and excluding faulty measurements, the integrity system shall generate a guaranteed protection level (PL), which bounds the estimation error (Groves 2013). Robust filtering of GNSS/INS measurements for accurate state estimation, together with a valid PL generation in the urban environment, remains under investigation and is targeted.

For the state estimation in changing environments, where the modeling of the satellite-measurement noise is often inadequate, a robust approach against environmental disturbances is crucial. Gehrt et al. (2020) have proposed a Bayesian-optimization-based adaptive parameterizing scheme within a Kalman filter, which chooses suitable parameters depending on the operating environment. The usage of this approach is limited to the designed environmental features. Instead of optimizing the parameter anytime and anywhere, robust estimation methods insensitive to environmental changes have been studied. As an alternative to Kalman filters, Simon (2006) has presented the $H_\infty$ filtering in detail, which still utilizes the conventional filter structure but shows significantly higher performance under system and environment uncertainties. In the context of GNSS/INS, Osman et al. (2019) have introduced a multi-sensor fusion localization approach using the EHF, where the wheel encoders, GPS, and visual odometry are fused within a loosely coupled structure. For tightly coupled GNSS/INS integration, Jwo and Chen (2003) have demonstrated the robustness of a hybrid $H_2/H_\infty$ filter against noise variations, assuming that the state transition is of constant-velocity kinematics. More recently, Smolyakov and Langley (2020) have proposed a novel algorithm combining $H_\infty$ filter and reinforcement learning for robust positioning solutions in urban environments. Our study focuses on the practical implementation of an EHF on a tightly coupled GNSS/INS system based on a modern strapdown architecture, emphasizing its robustness against various parameterizations and environmental circumstances.

For the PL generation, statistic approaches (Lee et al. 2018; Zhu et al. 2018) are widely applied and studied, assuming that the statistical distribution of the estimation error is known. Groves (2013) has indicated that such error distribution assumptions are difficult to validate. To release the PL generation from the statistical distribution conjecture, Combastel (2003) proposed set-membership approaches, which only require process and measurement errors to be bounded. In the context of GPS navigation, Dbouk and Schön (2018) compared various bounding methods, including zonotope, for GPS integrity and pointed out that the zonotope method describes the uncertainty propagation in a more meaningful manner. Dbouk and Schön (2020) further derived an outlier detection algorithm by applying zonotopes as uncertainty references for GPS integrity monitoring. In terms of zonotopes combined with an estimation observer, Zhang et al. (2020) have developed a robust and interval-based error estimation approach for descriptor systems and validated it in a simulation environment. Shetty and Gao (2021) recently conducted a non-linear stochastic reachability analysis to predict the state uncertainty bounds for GNSS-based unmanned aerial vehicle (UAV)-navigation, where probabilistic zonotopes, instead of conventional zonotopes, are applied as the set representation. The experimental validation was conducted in a simulated urban environment. More recently, Bhamidipati et al. (2022) applied zonotopes to represent GNSS shadows for 3D-map-aided GNSS localization. The feasibility and performance of using zonotope on GNSS/INS navigation, particularly on tightly coupled systems, remains to be researched.

Therefore, we propose a novel scheme combining the EHF and zonotope within the navigation system, involving a strapdown-based INS tightly coupled with GNSS pseudorange and deltarange observables. First, we use an EHF to estimate the states robustly against inappropriate parametrization of the GNSS measurement noise. Then, we apply zonotope to generate PL for the estimated vehicle states. The evaluation of the proposed methods uses a semi/sub-urban environment dataset from a commercial land vehicle. The EHF is compared with the extended Kalman filter (EKF) regarding the sensitivity of the state-estimation accuracy by grid searching a defined parameter space of GNSS measurement noise. Further, the state-estimation accuracy of the EHF and EKF are compared by applying the well-known sigma-ε model (Wieser et al. 2005; Breuer et al. 2016; Konrad et al. 2018) with the same GNSS parameters. Regarding the feasibility of applying zonotopes in real-time applications, we first evaluate the computational time of the 3D (for position PL) and 9D (for PL of the full vehicle states) zonotopes and their overestimation effects under various parameters. Finally, we apply the Stanford-ESA Integrity Diagram to validate the zonotope-based position-PL generation, checking the consistency of the estimated error bound with the ground truth. The result of zonotope-based PL estimation is briefly compared with the results in (Zhu et al. 2018), where the PL is estimated based on EKF measurement innovation by applying the sigma-ε model.

## Methodology

In the methodology part, we present the navigation scheme involving the EHF for the state estimation and zonotope for the PL generation. As shown in Fig. 1, the EHF is embedded into an existing GNSS/INS frame, introduced in the previous fundamental work (Konrad et al. 2018) from the same research group, including:

- The strapdown-based state and covariance prediction by applying the 100 Hz inertial measurement unit (IMU) measurements;
- The GNSS pre-processing, including the correction of various error terms;
- The handling of measurements at various rates with delay.

Additionally, GNSS fault detection and exclusion (FDE), proposed in previous works (Liu et al. 2019; Liu et al. 2021), is performed after the correction of GNSS observables, such that a set of fault-free GNSS measurements is fed into the EHF.

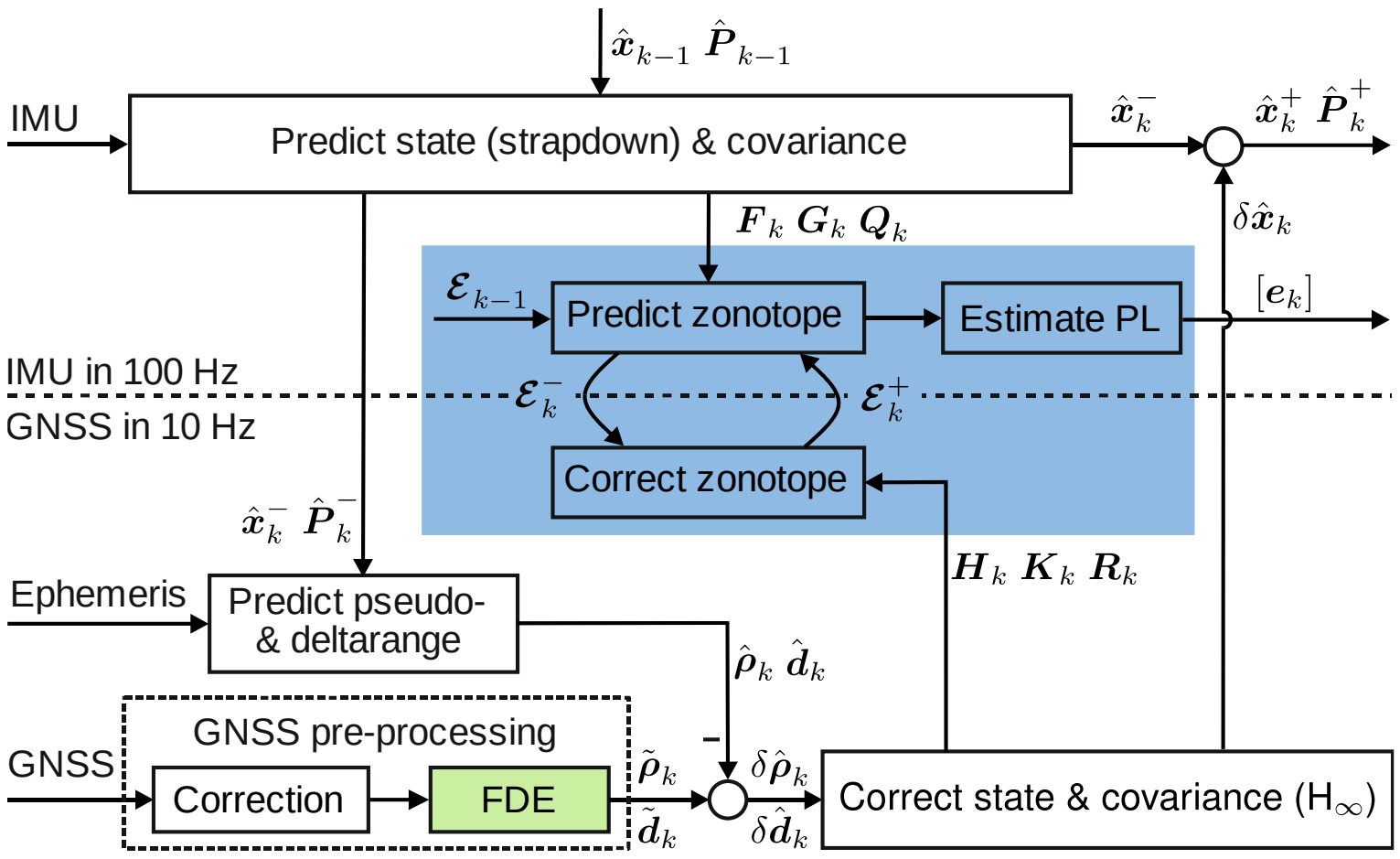

**Fig. 1** Overview of the designed tightly coupled navigation system, including transitions of the states, their covariances, and error bounds. The blue shadow marks the zonotope-relevant components. $k$ denotes the index of the time epoch.

The methodology of the EHF and zonotope-based PL generation is divided into two parts. The first part provides fundamental EHF and zonotope theories in the Appendix at the end of the manuscript. It is recommended to read the Appendix if the reader is unfamiliar with the corresponding theories. The second part explains their integration into the tightly coupled GNSS/INS in the rest of this section. A summary of the implementation in pseudocode form is given at the end of this section.

## State Estimation applying Extended H∞ Filter

Applying the EHF to estimate the vehicle's state, i.e., the three-dimensional position, velocity, and attitude, the following state vector $\boldsymbol{x}$ is designed:

$$\boldsymbol{x} = \left(\boldsymbol{p}_{eA}^{e,T}\ \boldsymbol{v}_{eA}^{n,T}\ \boldsymbol{q}_b^{n,T}\ \boldsymbol{b}_a^T\ \boldsymbol{b}_g^T\ c_b\ c_d\right)^T \tag{1}$$

where the index '$A$' stands for the antenna-body frame, '$b$' for the IMU-body frame, '$e$' for ECEF coordinates, and '$n$' for the navigation frame. $\boldsymbol{p}_{eA}^e \in \mathbb{R}^{3\times1}$ represents the position of the antenna body-frame origin in the earth-centered-earth-fixed (ECEF) frame. $\boldsymbol{v}_{eA}^n \in \mathbb{R}^{3\times1}$ is the velocity of the antenna-body-frame origin with respect to the ECEF frame, expressed in the navigation frame north-east-down (NED) coordinates. A quaternion $\boldsymbol{q}_b^n \in \mathbb{R}^{4\times1}$ for the alignment of the IMU body frame with the navigation frame is estimated as the expression of vehicle attitude.

Additionally, the accelerometer bias $\boldsymbol{b}_a$ and gyroscope bias $\boldsymbol{b}_g$, each of dimension $\mathbb{R}^{3\times 1}$, are part of the state vector, which is necessary for IMU integration. Further, the receiver clock bias $c_a$ and drift $c_d$ are estimated, which are essential for tightly coupled GNSS integration.

As shown in Fig. 1, the state vector $\boldsymbol{x}$ and its covariance $\boldsymbol{P}$ are propagated within a strapdown-based INS at 100 Hz. Konrad et al. (2018) have introduced the prediction of vehicle kinematics by applying the IMU accelerometer and gyroscope measurements and the prediction of IMU bias and receiver clock bias, summarized as (A4). Meanwhile, predicting the covariance by applying (A5) requires the system matrix $\boldsymbol{F}_k$ and the noise shaping matrix $\boldsymbol{G}_k$, the derivation of which is given in (Wendel 2009) or (Groves 2013).

Further, the predicted state vector and covariance are corrected by applying (A6)~(A8) whenever a new set of GNSS measurements is available. The GNSS measurements, i.e., pseudoranges and deltaranges, are introduced in the form of a measurement equation as (A2). By applying such nonlinear measurement equations in EHF, a linearization is required to calculate the measurement Jacobian matrix $\boldsymbol{H}_k$, which is sufficiently described in (Wendel 2009) or (Groves 2013).

*Measurement Model*

The measurement equation of pseudorange is given as follows:

$$\tilde{\rho} = \|\boldsymbol{p}_{es}^{e} - \boldsymbol{p}_{eA}^{e}\| + c_b + c_s + I_r + T_r + M_\rho + \nu_\rho \quad (2)$$

where $\tilde{\rho}$ is the measured scalar pseudorange of one satellite, $\boldsymbol{p}_{es}^{e}$ and $\boldsymbol{p}_{eA}^{e}$ represent the position of a satellite (denoted by '$s$') and the antenna in ECEF coordinates displayed in ECEF, respectively. The terms $c_b$, $c_s$, $I_r$, $T_r$, $M_\rho$, $\nu_\rho$ in (2) represent the receiver and satellite clock errors, the ionospheric error, the tropospheric error, the multi-path error, and the additive measurement noise, respectively. These error sources are corrected with differential correction data from a base station introduced in *Sec. Experiment Setup*.

The deltarange observation is calculated as:

$$\tilde{d} = (\boldsymbol{e}_{As}^{n})^T(\boldsymbol{v}_{es}^{n} - \boldsymbol{v}_{eA}^{n}) + c_d + \nu_d \quad (3)$$

where $\tilde{d}$ is the measured scalar relative velocity between the antenna and satellite, known as the deltarange. The vector $\boldsymbol{e}_{As}^{n}$ is the direction from the antenna to a satellite. The terms $\boldsymbol{v}_{es}^{n}$ and $\boldsymbol{v}_{eA}^{n}$ represent the velocity of the satellite and antenna with respect to the navigation frame, respectively. The terms $c_d$ and $\nu_d$ represent the clock drift and measurement noise, respectively.

In practice, the GNSS observables are measured in the antenna-body frame, while the strapdown algorithm propagates the system state vector in the IMU-body frame, shown in Fig. 1. In the navigation filter, this requires a transformation from the defined vehicle states $\boldsymbol{p}_{eA}^{e}$ and $\boldsymbol{v}_{eA}^{n}$ to the required states $\boldsymbol{p}_{eb}^{e}$ and $\boldsymbol{v}_{eb}^{n}$ in the IMU body-frame, shown as:

$$\boldsymbol{p}_{eb}^{e} = \boldsymbol{p}_{eA}^{e} - \boldsymbol{C}_{n}^{e}\boldsymbol{C}_{b}^{n}\boldsymbol{l}_{A}^{b} \quad (4)$$

$$\boldsymbol{v}_{eb}^{n} = \boldsymbol{v}_{eA}^{n} - \boldsymbol{C}_{b}^{n}\left(\boldsymbol{\omega}_{eb}^{b} \times \boldsymbol{l}_{A}^{b}\right) \quad (5)$$

where $\boldsymbol{C}_{n}^{e}$ and $\boldsymbol{C}_{b}^{n}$ denote the direction cosine matrix for the vector rotation among coordinate frames, derived from the estimated antenna position $\boldsymbol{p}_{eA}^{e}$ and the estimated attitude $\boldsymbol{q}_{b}^{n}$ (Groves 2013), respectively. The transformation uses the lever arm $\boldsymbol{l}_{A}^{b}$ of the antenna position in the IMU body-frame, measured and given in *Sec. Experiment Setup*, and considers the translation velocity of the antenna caused by its rotation $\boldsymbol{\omega}_{eb}^{b}$, derived from the IMU measurement and estimated IMU bias $\boldsymbol{b}_g$ (Wendel 2009).

(2) and (3) give the measurement equations of a single satellite and are applied to all observed satellites within the EHF as a measurement update step (Wendel 2009; Groves 2013; Konrad et al. 2018).

## Error-Bound Estimation using Zonotope

The core issue of the PL calculation is to bound the state estimation error $\boldsymbol{e_k}$. Analog to the filter state transition, the estimation of the state error bound $[\boldsymbol{e_k}]$ also contains a-priori propagation and a-posterior correction. The error dynamic of the EHF is given as follows:

$$\boldsymbol{e}_{k}^{-} = \boldsymbol{x}_{k} - \widehat{\boldsymbol{x}}_{k}^{-} = \boldsymbol{F}_{k-1}\boldsymbol{e}_{k-1}^{+} + \boldsymbol{G}_{k}\boldsymbol{w}_{k-1} \quad (6)$$

$$\boldsymbol{e}_{k}^{+} = \boldsymbol{x}_{k} - \widehat{\boldsymbol{x}}_{k}^{+} = (\boldsymbol{I} - \boldsymbol{K}_{k}\boldsymbol{H}_{k})\boldsymbol{e}_{k}^{-} - \boldsymbol{K}_{k}\boldsymbol{v}_{k} \quad (7)$$

where $\boldsymbol{e}_k^-$ and $\boldsymbol{e}_k^+$ represent the a-priori and a-posterior estimation error. $\hat{\boldsymbol{x}}_k^-$, $\hat{\boldsymbol{x}}_k^+$, and $\boldsymbol{K}_k$ are filter-relevant variables and defined in (A4)~(A8).

Assume that the initial estimation error $\boldsymbol{e}_0$, the process noise $\boldsymbol{w}_k$ and the measurement noise $\boldsymbol{v}_k$ are zero-mean and bounded by corresponding zonotopes:

$$\boldsymbol{e}_0 \in < \boldsymbol{0}, \boldsymbol{\mathcal{E}}_0 >, \boldsymbol{w}_k \in < \boldsymbol{0}, \boldsymbol{\mathcal{W}}_k >, \boldsymbol{v}_k \in < \boldsymbol{0}, \boldsymbol{\mathcal{V}}_k > \tag{8}$$

where $\boldsymbol{\mathcal{E}}_0$, $\boldsymbol{\mathcal{W}}_k$ and $\boldsymbol{\mathcal{V}}_k$ represent the generation matrix of corresponding zonotopes, defined in (A12). $\boldsymbol{\mathcal{E}}_0$ shall be initialized large enough, considering the large system uncertainties in the filter initialization stage. $\boldsymbol{\mathcal{W}}_k$ and $\boldsymbol{\mathcal{V}}_k$ are calculated as diagonal matrices with their diagonal elements equal to $n_\sigma$ times the standard deviation of process noise $\sigma_{w_k}$ and measurement noise $\sigma_{v_k}$, respectively. $n_\sigma$ is a pre-defined parameter corresponding to the confidence level of the resulting zonotope. Noteworthy, for the calculation of $\boldsymbol{\mathcal{W}}_k$ and $\boldsymbol{\mathcal{V}}_k$, we still apply the statistical distribution of the process and measurement noise, which is a zero-mean normal distribution. However, such distribution is not required by the implementing zonotope. Instead, zonotope only requires the lower and upper bound, which can be obtained from the sensor characteristics of the manufacturer or vehicle dynamic limitations. In general, the measurement boundaries are more convenient to obtain and validate in practice, which is one main advantage of set-membership approaches over statistic approaches.

After the zonotope is initialized with $\boldsymbol{\mathcal{E}}_0$, the a-priori and a-posteriori zonotope estimations are derived analogously to (6) and (7). The matrices multiplication and matrices addition in (6) and (7) correspond to linear image $\odot$ and Minkowski sum $\oplus$ in the zonotope calculation, explained with (A13) and (A14), respectively:

$$< \boldsymbol{c}_{e_k}^-, \boldsymbol{\mathcal{E}}_k^- > = \left(\boldsymbol{F}_{k-1} \odot < \boldsymbol{c}_{e_{k-1}}^+, \boldsymbol{\mathcal{E}}_{k-1}^+ >\right) \oplus (\boldsymbol{G}_{k-1} \odot < \boldsymbol{0}, \boldsymbol{\mathcal{W}}_{k-1} >) \tag{9}$$

$$< \boldsymbol{c}_{e_k}^+, \boldsymbol{\mathcal{E}}_k^+ > = \left((\boldsymbol{I} - \boldsymbol{K}_k \boldsymbol{H}_k) \odot < \boldsymbol{c}_{e_k}^-, \boldsymbol{\mathcal{E}}_k^- >\right) \oplus (\boldsymbol{K}_k \odot < \boldsymbol{0}, \boldsymbol{\mathcal{V}}_k >) \tag{10}$$

where $\boldsymbol{c}_{e_k}^-$, $\boldsymbol{c}_{e_k}^+$, $\boldsymbol{\mathcal{E}}_k^-$, $\boldsymbol{\mathcal{E}}_k^+$ are the center and generation matrix of the a-priori and a-posteriori estimated bounding zonotope for the state estimation error $\boldsymbol{e_k}$. Since $< \boldsymbol{c}_{e_k}^-, \boldsymbol{\mathcal{E}}_k^- >$ and $< \boldsymbol{c}_{e_k}^+, \boldsymbol{\mathcal{E}}_k^+ >$ are iteratively updated using the zero-mean zonotopes $< \boldsymbol{0}, \boldsymbol{\mathcal{E}}_0 >$, $< \boldsymbol{0}, \boldsymbol{\mathcal{W}}_k >$, $< \boldsymbol{0}, \boldsymbol{\mathcal{V}}_k >$, the center of the resulting zonotope $\boldsymbol{c}_{e_k}^-$ and $\boldsymbol{c}_{e_k}^+$ shall be zero, which means

$$\boldsymbol{e}_k \in < \boldsymbol{0}, \boldsymbol{\mathcal{E}}_k >, \ \forall k \in \{1, \cdots, N\} \tag{11}$$

Therefore, only the estimation of the generation matrix $\boldsymbol{\mathcal{E}}_k^-$ and $\boldsymbol{\mathcal{E}}_k^+$ is necessary. (9) and (10) result in the following zonotope propagation and update steps:

$$\boldsymbol{\mathcal{E}}_k^- = \mathcal{R}([\boldsymbol{F}_{k-1}\boldsymbol{\mathcal{E}}_{k-1}^+ \quad \boldsymbol{G}_{k-1}\boldsymbol{\mathcal{W}}_k\ ]) \tag{12}$$

$$\boldsymbol{\mathcal{E}}_k^+ = \mathcal{R}([(\boldsymbol{I} - \boldsymbol{K}_k\boldsymbol{H}_k)\boldsymbol{\mathcal{E}}_k^- \quad \boldsymbol{K}_k\boldsymbol{\mathcal{V}}_k]) \tag{13}$$

Each zonotope estimation carries out a zonotope reduction, explained with (A16) and denoted as '$\mathcal{R}$'. The dimension of the resulting zonotope is specified by a user-defined reduction order $q$ $(n \leq q \leq m)$, namely $\mathcal{R}(\boldsymbol{\mathcal{H}}) \in \mathbb{R}^{n\times q}$. Without applying the zonotope reduction, the dimension of the generation matrices $\boldsymbol{\mathcal{E}}_k^-$ and $\boldsymbol{\mathcal{E}}_k^+$ increases without limit along the operation time, which is not affordable for computation units.

In the EHF, the state vector $\boldsymbol{x}$ contains 18 states, including the quaternion $\boldsymbol{q}_b^n \in \mathbb{R}^{4\times 1}$. By the estimation of the state error covariance $\boldsymbol{P}$ , the error of the orientation vector $\mathbb{R}^{3\times 1}$ is considered (Wendel 2009), which makes $\boldsymbol{P} \in \mathbb{R}^{17\times 17}$. Therefore, the error space of the EHF is 17-dimensional. It results in the generation matrix for the EHF $\boldsymbol{\mathcal{E}} \in \mathbb{R}^{17\times q}$. However, in practice, the PLs of some states could be undesired, e.g., PLs of the IMU bias. Suppose that the sequence of the index of the desired PLs in the EHF's error space is given as

$$\boldsymbol{\tau} = \{\tau_1,\ \tau_2,\ \cdots,\ \tau_n\} \tag{14}$$

where $n$ is the dimension of the desired zonotope. A selection matrix $\boldsymbol{\Gamma} \in \mathbb{R}^{n\times 17}$ can be applied to the generation matrix calculation to achieve the state selection. $\Gamma_{\boldsymbol{ij}}$ denotes the element of the i$^{\text{th}}$ column and j$^{\text{th}}$ raw in $\boldsymbol{\Gamma}$ and is given as:

$$\Gamma_{ij} = \begin{cases} 1, & \text{for } i = j\ \in \boldsymbol{\tau} \\ 0, & \text{otherweise} \end{cases},\ \forall i \in \{1,\ 2,\ \cdots,\ n\},\ \forall j \in \{1,\ 2,\ \cdots,\ 17\} \tag{15}$$

The calculation of the zonotope $< \boldsymbol{0}, \boldsymbol{\mathcal{E}}_{r,k} >$ to bound the selected state errors is given as follows:

$$\boldsymbol{\mathcal{E}}_{r,k}^- = \mathcal{R}\left(\left[\boldsymbol{\Gamma}\boldsymbol{F}_{k-1}\boldsymbol{\Gamma}^T\boldsymbol{\mathcal{E}}_{r,k-1}^+ \quad \boldsymbol{\Gamma}(\boldsymbol{G}_{k-1}\boldsymbol{\mathcal{W}}_{k-1})\right]\right) \tag{16}$$

$$\boldsymbol{\mathcal{E}}_{r,k}^+ = \mathcal{R}\left(\left[\boldsymbol{\Gamma}(\boldsymbol{I} - \boldsymbol{K}_k\boldsymbol{H}_k)\boldsymbol{\Gamma}^T\boldsymbol{\mathcal{E}}_{r,k}^- \quad \boldsymbol{\Gamma}(\boldsymbol{K}_k\boldsymbol{\mathcal{V}}_k)\right]\right) \tag{17}$$

In practice, the error bound of the position, velocity, and orientation could be of interest, which can be estimated by the nine-dimensional zonotope given as $\boldsymbol{\mathcal{E}}_{\boldsymbol{r},\boldsymbol{9D}} \in \mathbb{R}^{9\times q}$. The zonotope could be further reduced to three-dimensional $\boldsymbol{\mathcal{E}}_{\boldsymbol{r},\boldsymbol{3D}} \in \mathbb{R}^{3\times q}$ if only the PL of the position is required. In the present work, the three- and nine-dimensional zonotopes are implemented and discussed, which are achieved by selection matrices $\mathbf{\Gamma}_{\mathbf{3}D} \in \mathbb{R}^{3\times 17}$ and $\mathbf{\Gamma}_{\mathbf{9}D} \in \mathbb{R}^{9\times 17}$:

$$\boldsymbol{\tau}_{\mathbf{3}D} = \{1,\ 2,\ 3\},\ \mathbf{\Gamma}_{\mathbf{3}D} = [\boldsymbol{I}_{\mathbf{3}\times\mathbf{3}} \quad \mathbf{0}_{\mathbf{3}\times\mathbf{14}}] \tag{18}$$

$$\boldsymbol{\tau}_{\mathbf{9}D} = \{1,\ 2,\ \cdots,\ 9\},\ \mathbf{\Gamma}_{\mathbf{9}D} = [\boldsymbol{I}_{\mathbf{9}\times\mathbf{9}} \quad \mathbf{0}_{\mathbf{9}\times\mathbf{8}}] \tag{19}$$

These two specific zonotopes will be evaluated in the experiment section.

## Protection-Level Generation and Integrity Risk

In general, the objective of estimating the PL is to find the boundary of estimation error under an integrity risk:

$$\boldsymbol{PL}_k = \sup\{[\boldsymbol{e}_k] | P(\hat{\boldsymbol{x}}_k - \boldsymbol{x}_k \in [\boldsymbol{e}_k]) \leq 1 - P_{IR}\} \tag{20}$$

where $\boldsymbol{x}_k$ is the unknown ground truth of the state estimate $\hat{\boldsymbol{x}}_k$. $P_{IR}$ denotes the probability of integrity risk. $\boldsymbol{PL}_k$ is the vector calculated for all desired states and corresponds to the dimension of $\boldsymbol{\tau}$.

Assuming that (8) is satisfied, (11) indicates that the estimation error must be within the zonotope given as $< \mathbf{0}, \boldsymbol{\mathcal{E}}_k >$ and, therefore, must be within its interval hull. Thus, we estimate the desired protection level as the interval hull box of the resulting reduced zonotope $< \mathbf{0}, \boldsymbol{\mathcal{E}}_{r,k} >$ from (16) and (17) by applying (A15). In practice, specific PLs could be desired, e.g., horizontal PL (HPL) and vertical PL (VPL) for positioning. The HPL is calculated as the diagonal distance of the interval hull box corresponding to positioning error in the north and east direction, while the VPL is the interval hull of positioning error in the down direction.

The risk of initializing the zonotopes $< \mathbf{0}, \boldsymbol{\mathcal{E}}_0 >$, $< \mathbf{0}, \boldsymbol{\mathcal{W}}_k >$, and $< \mathbf{0}, \boldsymbol{\mathcal{V}}_k >$ can be quantified with the pre-defined parameter $n_\sigma$. However, the zonotope operations sustainably change the integrity risk, such as Minkowski sum, or even deterministically reduce the integrity risk, such as zonotope reduction and interval hull generation. Since the zonotope is independent of the statistical distributions after being initialized with (8), the resulting integrity risk $P_{IR}$ is hard to analyze in theory. We evaluate the integrity risk of zonotope-based protection-level generation by applying the Stanford-ESA diagram with recorded real-world data.

Summary of the Implementation Scheme

The integration of the zonotope into the filter structure is summarized as Alg. 1. It should be noted that Alg. 1 is a general expression that does not require the applied filter to be the EHF. Any other form of filter, e.g., the EKF or the unscented Kalman filter (UKF), could also be applied as long as its error dynamic is known. Applying another filter requires adjusting (6), (7), (16), and (17) according to the filter dynamic.

---

**Algorithm 1** EHF Extended with Zonotope

---

Initialization:

EHF: $\boldsymbol{x}_0, \boldsymbol{P}_0$; Zonotope: $\boldsymbol{\mathcal{E}}_0$

Propagation:

EHF: $\hat{\boldsymbol{x}}_{k-1} \rightarrow \hat{\boldsymbol{x}}_k^-, \hat{\boldsymbol{P}}_{k-1} \rightarrow \hat{\boldsymbol{P}}_k$ (A4)(A5)

Zonotope: $\boldsymbol{\mathcal{E}}_{r,k-1} \rightarrow \boldsymbol{\mathcal{E}}_{r,k}^-$ (16)

Measurement Update:

**if** new measurements are available, **then**

EHF: $\hat{\boldsymbol{P}}_k^- \rightarrow \hat{\boldsymbol{P}}_k^+, \hat{\boldsymbol{x}}_k^- \rightarrow \hat{\boldsymbol{x}}_k^+$ (A6)(A7)(A8)

Zonotope: $\boldsymbol{\mathcal{E}}_{r,k}^- \rightarrow \boldsymbol{\mathcal{E}}_{r,k}^+$ (17)

**end if**

Output Generation:

**if** new measurements are available, **then**

EHF: $\hat{\boldsymbol{P}}_k = \hat{\boldsymbol{P}}_k^+, \hat{\boldsymbol{x}}_k = \hat{\boldsymbol{x}}_k^+$; Zonotope: $\boldsymbol{\mathcal{E}}_k = \boldsymbol{\mathcal{E}}_k^+$

**else**

EHF: $\hat{\boldsymbol{P}}_k = \hat{\boldsymbol{P}}_k^-, \hat{\boldsymbol{x}}_k = \hat{\boldsymbol{x}}_k^-$; Zonotope: $\boldsymbol{\mathcal{E}}_k = \boldsymbol{\mathcal{E}}_k^-$

**end if**

Protection Level: $\boldsymbol{\mathcal{E}}_{r,k} \rightarrow [\boldsymbol{e}_{r,k}]$ (A15)

---

## Experimental Evaluation

In this section, we first introduce the hardware and software setup for data recording, followed by the parameterization of the variables mentioned in the methodology. We designed and summarized three experiments using the dataset, providing their results in a separate section.

### Experiment Setup

The designed approach is validated in a post-processing environment using the data recorded on a commercial land vehicle in a real-world test campaign. Fig. 3 shows the experimental vehicle with the measurement setups for real-time data recording. The lever arm of the antenna position in the IMU body-frame $\boldsymbol{l}_A^b$ is measured as:

$$\boldsymbol{l}_A^b = (0 \quad 0 \quad -0.1131\ \mathrm{m})^T \tag{21}$$

which is applied in (4) and (5) for the lever arm compensation.

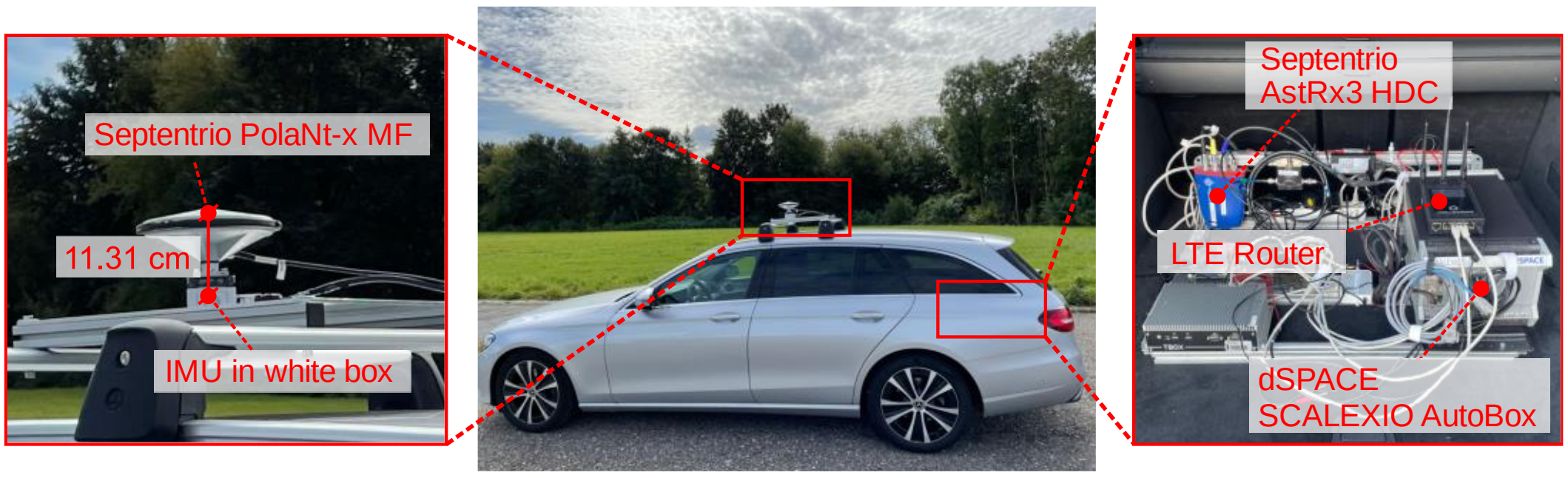

**Fig. 2** Measurement setup equipped on a car: the antenna and IMU were mounted on the roof (left); the rest of the hardware was placed in the trunk (right).

Table 1 gives a list of the hardware and the corresponding parametrization. Specifically, the parameterization of the measurement noise ($\nu_\rho$ and $\nu_d$ ) variance matrix $\boldsymbol{R}$ for all GNSS observables is crucial. Breuer et al. (2016) and Konrad et al. (2018) apply the sigma-ε model for this purpose, which calculates the standard deviation for each pseudorange and deltarange as:

$$\sigma_\rho^2 = C_\rho^2 \cdot 10^{-\frac{\mathrm{C/N_0}}{10}}, \; \sigma_d^2 = C_d^2 \cdot 10^{-\frac{\mathrm{C/N_0}}{10}} \tag{22}$$

where $\mathrm{C/N_0}$ is the carrier-to-noise ratio in dB-Hz and provided by the GNSS receiver. $\mathrm{C/N_0}$ varies among satellites and with time. $C_\rho$ and $C_d$ are tuning parameters defined for the pseudorange and deltarange measurements. The choice of $C_\rho$ and $C_d$ is evaluated and discussed in the next section.

**Table 1** List of the applied hardware and the specifications. The PPS is used for measurement delay compensation (Konrad et al. 2018).

| Hardware | Function/measurement (rate) | Parametrization |
|---|---|---|
| dSPACE SCALEXIO Autobox | Real-time data recording (1000 Hz) | N/A |
| LORD MicroStrain 3DM®-GX5-25 IMU | Acceleration (100 Hz)<br>Angular rate (100 Hz) | Process noise variance $\boldsymbol{Q}$: Allan variance analysis in (Breuer et al. 2016) |
| Septentrio AstRx3 HDC GNSS receiver | Pseudorange (10 Hz)<br>Deltarange (10 Hz) | Measurement noise variance $\boldsymbol{R}$: sigma-ε model in (22) |
| | Pulse per second (PPS, 1 Hz)<br>Real-time-kinematic ground-truth solution (RTK, 10 Hz) | N/A |
| Raspberry Pi 4 | Differential correction data from SAPOS® NRW (1 Hz) | N/A |

The initial estimation error covariance matrix of the EHF $\boldsymbol{P}_0$ is calculated by the initial standard deviation $\boldsymbol{\sigma}_0$ given in Table 2.

**Table 2** Standard deviations of the initial states in the EHF

| **state** | pos. e/n | pos. d | velocity | orientation |
|---|---|---|---|---|
| $\sigma_0$ | 5 m | 10 m | 1 m/s | 5° |

| **state** | acc. bias | gyro. bias | clock drift | clock bias |
|---|---|---|---|---|
| $\sigma_0$ | 0.1 m/s$^2$ | 0.01°/s | 1000 m | 10 m/s |

The reduction order parameter $q$ of zonotope is introduced in (A16) and is discussed with experimental results later in this section. Besides, $n_\sigma$ is set to 3 for the zonotope initialization.

## Experiment Overview

In this section, the test drive of the experimental evaluation is described, followed by an overview of the designed experiments. Fig. 3 shows the bird-eye view of the RTK reference trajectory on the Google-Map-satellite-view layer. The testing area locates in Aachen, Germany, and is a mix of boulevard, semi-urban, sub-urban, and highway environments. The detailed testing area division refers to (Nitsch et al. 2021), where the test drive was carried out in the same area but applied to another measurement setup.

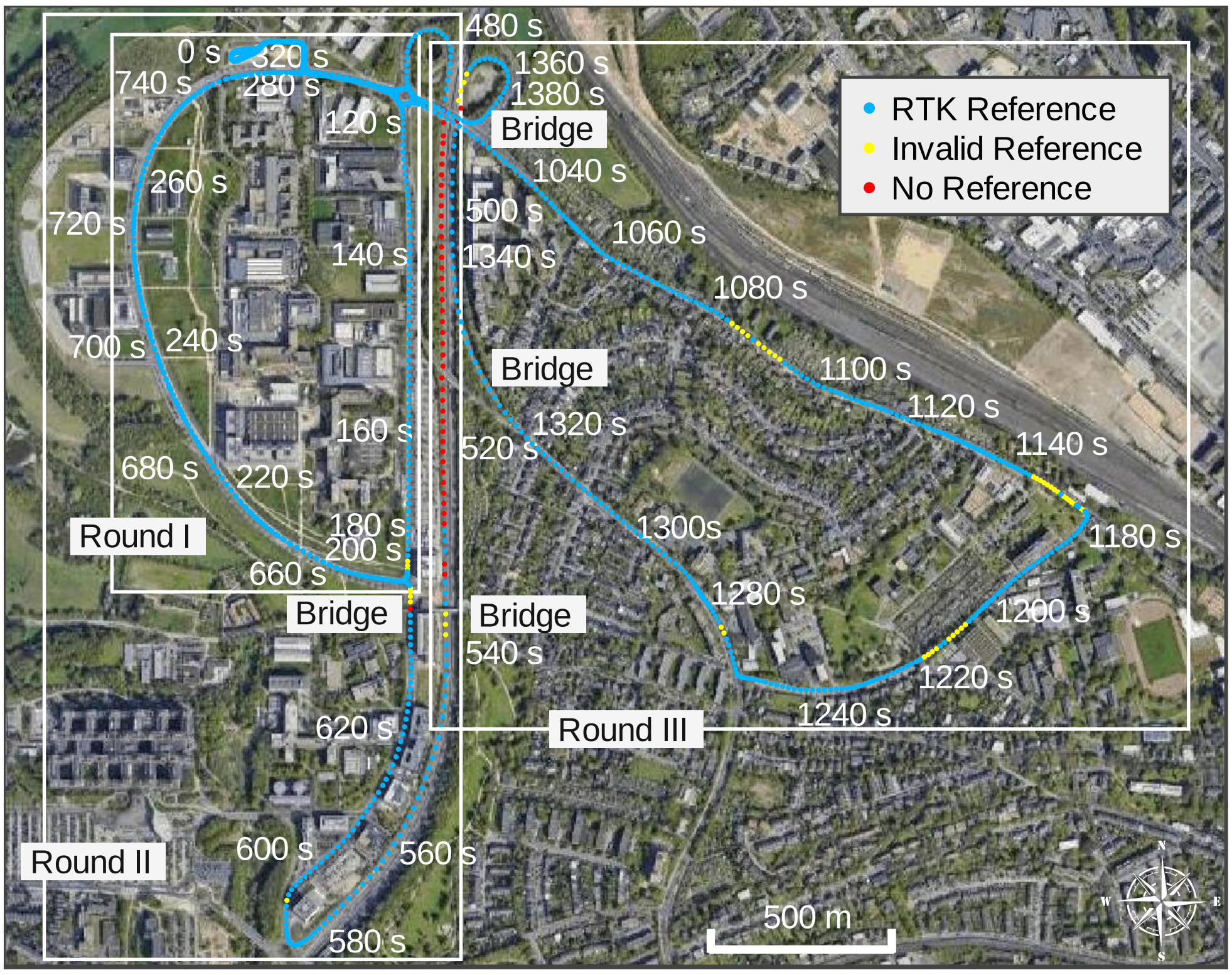

**Fig. 3** Bird-eye view of the reference trajectory: valid RTK reference position in blue; invalid non-RTK reference position in yellow; unavailable reference position in red, where the positioning solution is estimated by the navigation filter. Map data ©2022 GeoBasis-DE/BKG (©2009), Google.

The entire test drive takes 1514 seconds and is divided into three rounds, the overview of which is given in Table 3.

**Table 3** Description of the three test rounds

| | Round I | Round II | Round III |
|---|---|---|---|
| Time | 0 - 360 s | 360 - 900 s | 900 - 1514 s |
| Location | circle inner left | circle outer left | circle right |
| Direction | clockwise | clockwise | clockwise |
| Environment | boulevard, semi-urban | boulevard, semi-urban, highway | boulevard, sub-urban, highway |
| Signal Disturbance | low | high | high |

Combining Fig. 3, Fig. 4, and Table 3, in Round I, the vehicle drives at a speed of up to 50 km/h in an interference-lack environment, where the number of available satellites is stable, and the $C/N_0$ plot indicates that the signal quality is high and stable. Round II and III are much more challenging for satellite navigation, with high-speed tracks up to 70 km/h, and the vehicle drives under bridges, as marked in Fig. 3. The changing number of visible satellites and their unstable $C/N_0$ even caused failed positioning from the reference receiver for a long period of around 500 seconds.

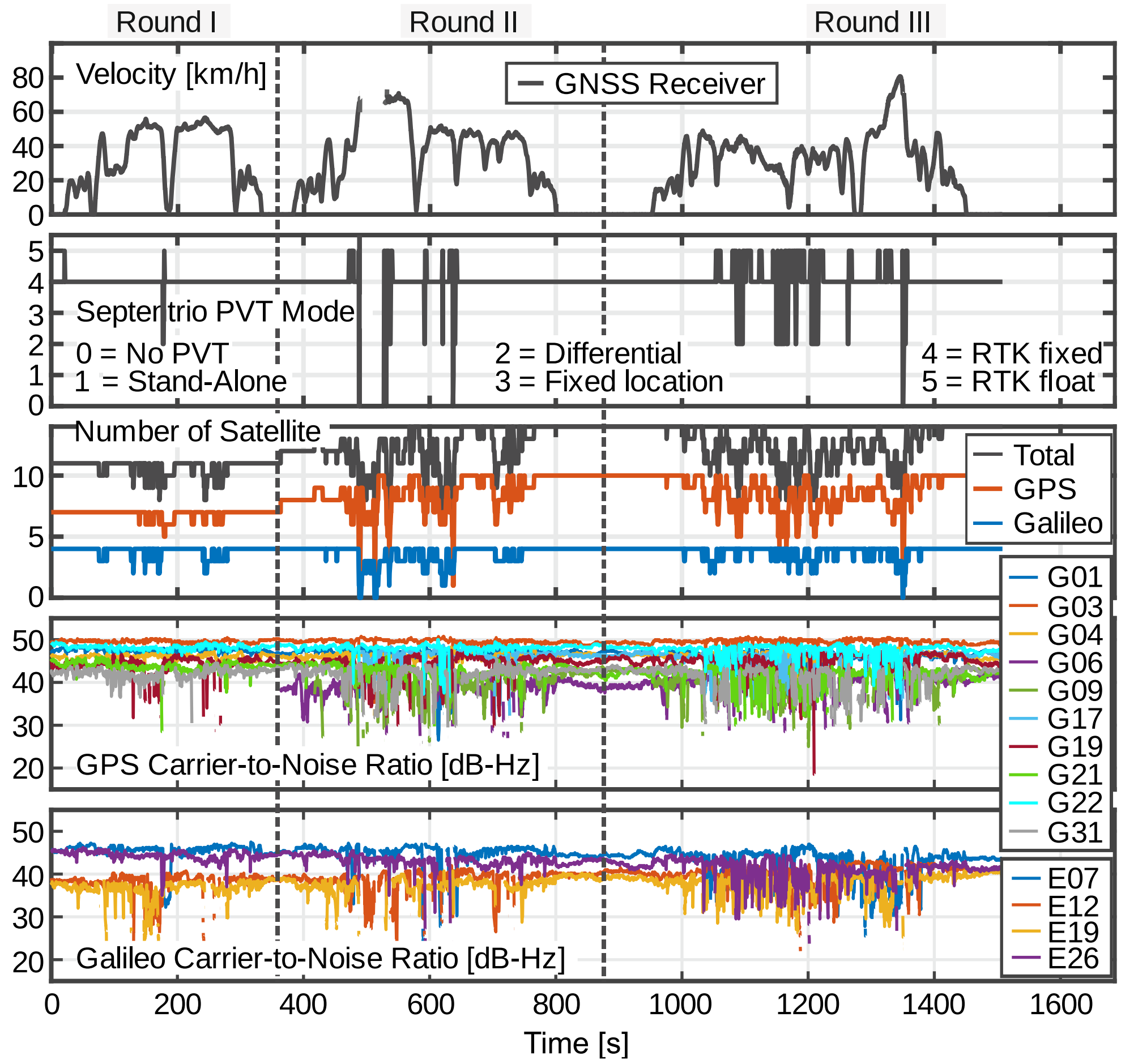

**Fig. 4** Receiver data during the test drive: the absolute velocity estimated by the GNSS receiver, the PVT mode of the receiver where only RTK fix and float can be considered as valid reference, the number of available satellites during the drive, and the carrier-to-noise ratio of the tracked GPS and Galileo signals.

Considering the characteristics of the current dataset, we split the entire test drive into a dataset for parametrization (Round I) and a dataset for validation (Round II and III). And following experiments are designed:

- *Sec. Sensitivity Analysis of Estimation Error vs. $C_\rho$ and $C_d$*: Here, we analyze the sensitivity of the state estimation accuracy from the EHF and EKF by applying a grid search upon the parameter set $C_\rho$ and $C_d$. This experiment uses the dataset **Round I** and determines the $C_\rho$ and $C_d$ that are applied in the following experiments.
- *Sec. Comparison between EHF and EKF regarding State-Estimation Accuracy*: Here, we evaluate the state estimation accuracy of EHF and EKF using the datasets **Round II** and **III**, with the determined parameters from Round I. The idea is to evaluate the robustness against sub-optimal parameter settings since the applied parameter set is determined in a different environment, namely Round I, rather than its actual operation environment, namely Round II and III, which contains much more signal disturbances.

- *Sec. Validation of PL Generation using Zonotope*: Here, we discuss the choice of reduction order $q$ by applying three and nine-dimensional zonotopes and validate the PL estimation by applying the Stanford-ESA Integrity Diagram using the validation dataset **Round II** and **III**.

Sensitivity Analysis of Estimation Error vs. $C_\rho$ and $C_d$

This section analyses and compares the sensitivity of the state-estimation accuracy from EHF and EKF against the parameter set $C_\rho$ and $C_d$. A grid search is performed within the parameter range $C_\rho \in [10, 4000]\ \mathrm{m}/\sqrt{\mathrm{Hz}}$ and $C_\mathrm{d} \in [0.25, 192]\ \mathrm{m}/\left(\mathrm{s}\cdot\sqrt{\mathrm{Hz}}\right)$. Considering the C/N0 values plotted in Fig. 4, which are within the range of $\mathrm{C/N0} \in [20, 50]$ dB-Hz, the range of the standard deviation of pseudorange and deltarange observables can be estimated by applying the sigma-ε model given in (22), and summarized in Table 4.

**Table 4** Standard deviation of the pseudorange and deltarange estimated by the sigma-ε model. * refers to the parameter choice for the following evaluations by applying (22) as a cost function.

| Estimated Sigma | | Pseudorange $C_\rho$ $\mathrm{m}/\sqrt{\mathrm{Hz}}$ | | | Deltarange $C_d$ $\mathrm{m}/\left(\mathrm{s}\cdot\sqrt{\mathrm{Hz}}\right)$ | | |
|---|---|---|---|---|---|---|---|
| | | 10 | 100* | 4000 | 0.25 | 1* | 192 |
| $\mathrm{C/N_0}$ | 50 | $3.2 \times 10^{-2}$ | 0.32 | 12.7 | $7.9 \times 10^{-4}$ | $3.2 \times 10^{-3}$ | 0.61 |
| dB-Hz | 20 | 1 | 10 | 400 | $2.5 \times 10^{-2}$ | 0.1 | 19.2 |

The grid search results are shown in Fig. 5. Comparing the subfigure top left and right, the minimum average 3D positioning error of both filters is approximately 0.62 m, indicating that the EHF has no advantage over the EKF under optimal parameter settings. However, the positioning accuracy of the EHF is less sensitive to the parameters (varies between 0.62 m to 1 m) than the EKF (varies between 0.62 m to 2 m). Comparing the subfigure bottom left and right, the EHF is slightly less sensitive to the parameters regarding the velocity-estimation accuracy. It should be noted that these results are against the statement in (Simon 2006, pp. 368): ‘the filter (EHF) performance is more sensitive to the design parameters. The reason is that, instead of the numerical example in (Simon 2006), the navigation filter has much larger system uncertainties, such as GNSS signal disturbances. Under poor parametrizations, the unmodeled measurement error dominates the filter’s performance. In this case, the EHF outperforms the EKF due to its filter robustness against system disturbances.

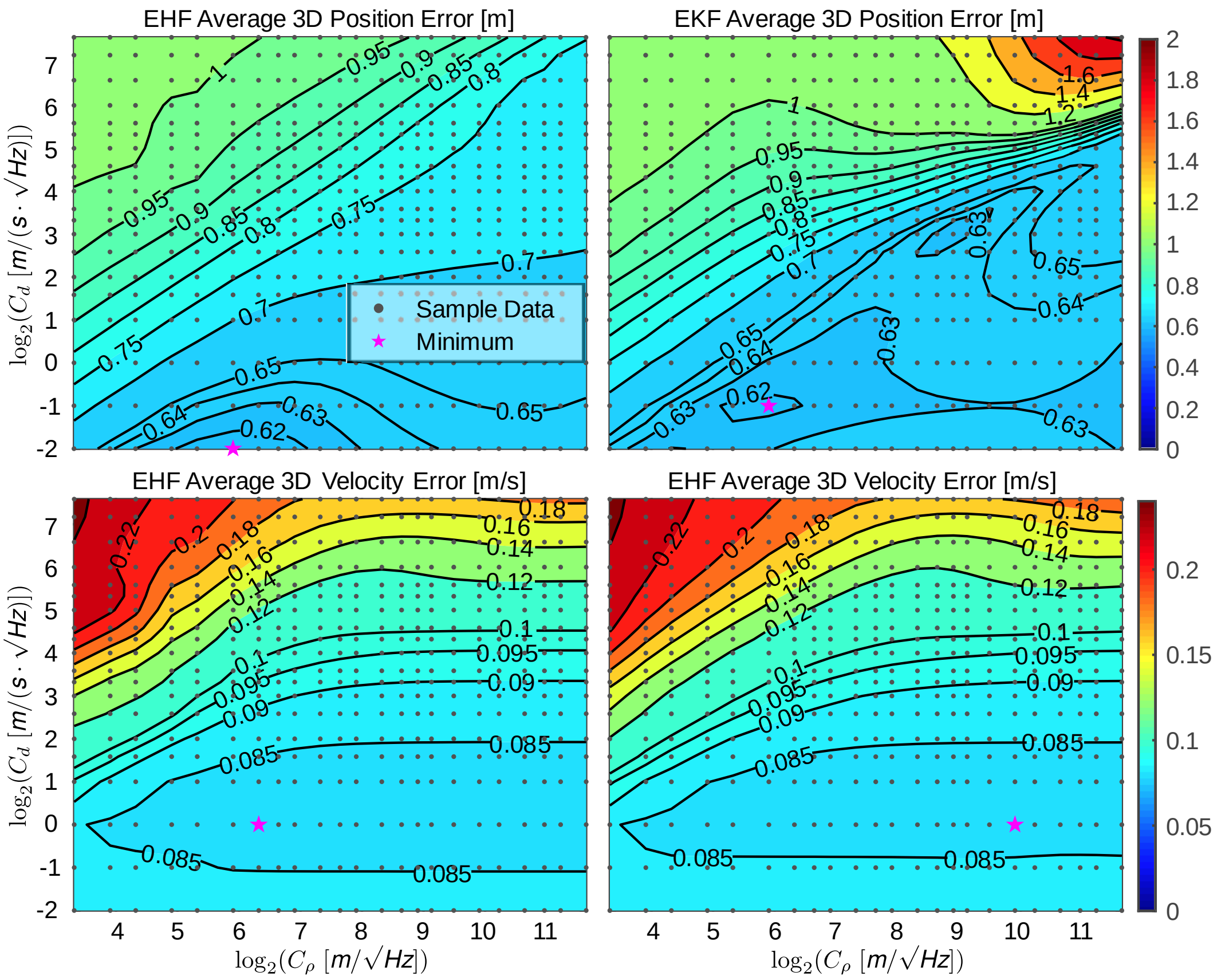

**Fig. 5** Average position and velocity-estimation accuracy of the EHF and EKF under various $C_\rho$ and $C_d$ applying the data from Round I. Each sample data is an individual simulation, and the contour plots are generated with the MATLAB® contourf.m function with default data interpolation. Note that both the x- and y-axes are logarithmic representations.

As mentioned in the experiment overview, a parameter set of $C_\rho$ and $C_d$ shall be chosen and used for the following evaluations. It is noticeable that the minimum values are different in each subfigure in Fig. 5, meaning that there are no optimal parameter settings for both filters. Considering that the evaluation in the next section aims at proving the robustness of the EHF under non-optimal conditions, we chose the parameter set, therefore, for the best benefit of the EKF, applying the following cost function:

$$\text{weighted 3D error} = \frac{\text{mean 3D position error}}{max(\text{ mean 3D position error })} + \frac{\text{mean 3D velocity error}}{max(\text{mean 3D velocity error })} \quad (23)$$

With (23), the 3D position and velocity error under each parameter setting are normalized by the maximum error of all parameter combinations and equally weighted. Applying (23), the parameter set is chosen as $C_\rho = 100\ \mathrm{m}/\sqrt{\mathrm{Hz}}$ and $C_d = 1\ \mathrm{m}/\left(\mathrm{s} \cdot \sqrt{\mathrm{Hz}}\right)$. The corresponding standard-deviation ranges estimated by the sigma-ε model are given in Table 4.

## Comparison between EHF and EKF regarding State-Estimation Accuracy

This section evaluates the state-estimation accuracy of the EHF and EKF using the validation data from Round II and III, applying the parameter set $C_\rho = 100\ \mathrm{m}/\sqrt{\mathrm{Hz}}$ and $C_d = 1\ \mathrm{m}/\left(\mathrm{s} \cdot \sqrt{\mathrm{Hz}}\right)$, which is optimized using data from Round I for the EKF. Fig. 6 shows the experimental results, and Table 5 shows various metrics of positioning accuracy.

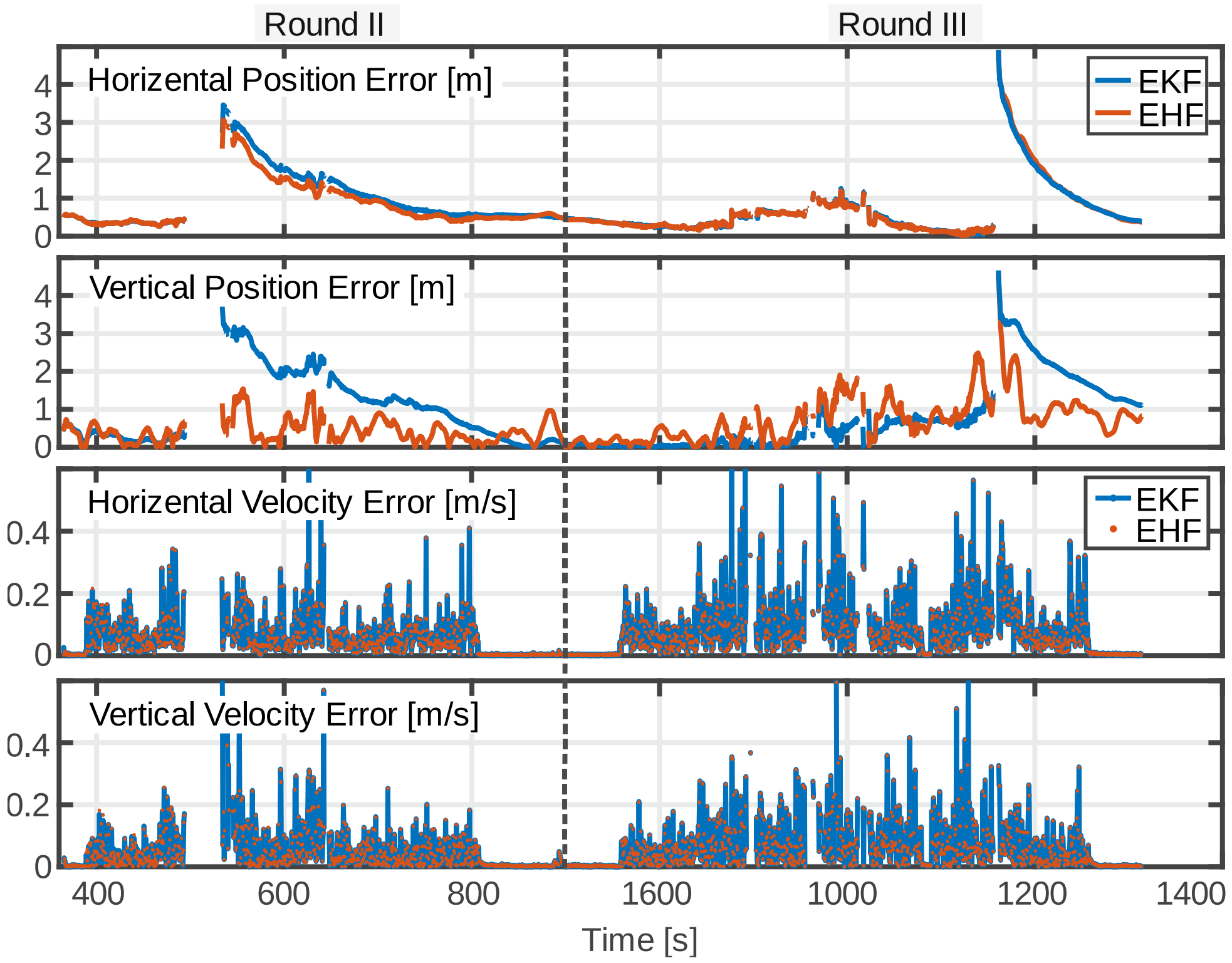

**Fig. 6** Comparison of the estimation accuracy of the EHF and EKF: horizontal and vertical position error in meter; horizontal and vertical velocity error in meter per second

The first subfigure shows that the EHF has a slight advantage over the EKF regarding horizontal positioning accuracy, the same as in Table 5. However, under extreme disturbances, e.g., at around 500 and 1350 seconds when the vehicle drives under a bridge and then directly into environments that lack satellite signals, both the EHF and EKF experience horizontal error up to meters.

**Table 5** Accuracy indicators of horizontal and vertical positioning error of EHF and EKF. "Std." stands for standard deviation, and "RMS" for root mean square.

| Positioning error [m] | Horizontal | | | Vertical | | |
|---|---|---|---|---|---|---|
| | Mean | Std. | RMS | Mean | Std. | RMS |
| EHF | 0.730 | 0.668 | 0.989 | 0.582 | 0.505 | 0.771 |
| EKF | 0.779 | 0.707 | 1.052 | 0.878 | 0.897 | 1.255 |

Meanwhile, the vertical error comparison in the second subfigure and Table 5 indicates a more accurate estimation from the EHF than from the EKF. Under extreme disturbances, the EHF maintains a vertical error of under 2 m, while the EKF has a vertical error of up to 4 m. The results of the horizontal and vertical position errors show that EHF is more robust against environmental changes and signal disturbances by applying the same parameters.

Noteworthy, applying the EKF, the horizontal estimation error is smaller than the vertical one, which is consistent with expectations: During the drive, the average horizontal and vertical dilution of precision (DOP) is 0.862 and 1.707, respectively. This results from the satellite-constellation limitation that the satellites distribute only above the horizon (Kaplan and Hegarty 2005, pp. 328-332). With the given DOPs, the vertical positioning uncertainty is expected to be twice as large as the one in the horizontal direction if a least square estimator was applied. However, applying the EHF, the horizontal estimation error is surprisingly smaller than the vertical one, which further proves the robustness and advantage of EHF against non-optimal operation conditions, i.e., non-optimal satellite constellation.

Finally, the third and fourth subfigures show no significant difference between EHF and EKF regarding the accuracy of velocity estimation, which is similar to the results in Fig. 5. Due to the lack of accurate orientation reference, the accuracy of the orientation estimation is not included.

More experiment results can be found in the supplementary material, where the EHF outweighed the EKF under various parametrizations of GNSS measurement noise, not only in vertical but also in horizontal positioning. Under extreme system disturbances, the EKF estimate diverges from the ground truth, while the EHF estimation error maintains limited, proving the EHF's theoretical strength.

## Validation of PL Generation using Zonotope

This section first discusses the choice of the reduction order $q$ and its influence on the zonotope estimation. It then validates the correctness of the PL estimation by applying the Stanford-ESA Integrity Diagram. Noteworthy, all PLs generated by zonotopes in this section are $3\sigma$-boundaries, corresponding to an interval integrity risk of 0.27% by initialization. It should be emphasized that the overall integrity risk shall be much lower, as discussed with (20).

Fig. 7 shows the average width of the estimated PL by applying three- and nine-dimensional zonotopes under various reduction order $q$ within the EHF. We conducted the post-processing using MATLAB & Simulink platform on a PC equipped with an Intel® Core (TM) i7-11700K @ 3.60GHz CPU. The results in Fig. 7 include the run time of the EHF together with the zonotope computed on a single CPU core.

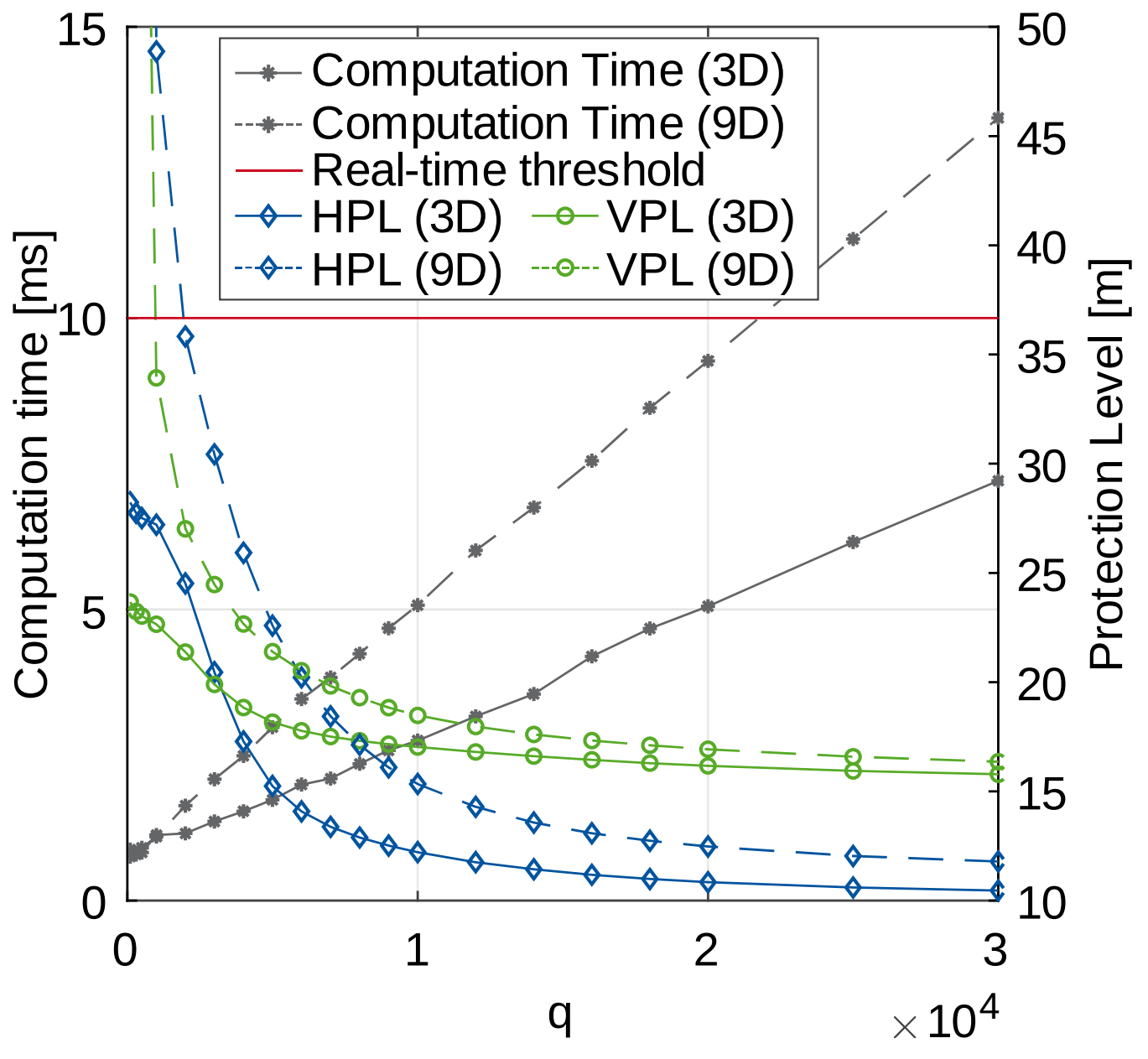

**Fig. 7** Computation time and mean value of the generated HPL and VPL for the position estimation using 3D and 9D zonotopes with various reduction order $q$.

As discussed in Fig. A1, the smaller $q$ indicates a larger overestimation of the zonotope and a smaller computational load. This is verified by the results in Fig. 7, where the following observations should be stressed:

- The computational time of the zonotopes increases quasi-proportionally to the reduction order $q$.
- The computational time of the zonotopes is not proportional to the dimension of the zonotopes. With the same reduction order $q = 20000$, the run time of the 9D zonotope is only 1.8 times that of the 3D zonotope.
- Applying high-dimensional zonotopes makes the overestimation effect more severe, especially with small reduction orders. By reduction order $q = 100$, 300, and 500, the average HPL (9D) is approximately $8.7 \cdot 10^{27}$, $1.7 \cdot 10^{4}$, and 137 m, respectively. For better visualization, they are not shown in Fig. 7.

- With small reduction orders, the estimated HPL is larger than the VPL because zonotope's overestimation effect dominates the PL generation. The calculation of HPL regards two states (position in north and east direction), which is affected by the overestimation more severely than VPL, which only regards one state (position in down direction). The overestimation effect becomes less dominant with the increasing reduction order, and the measurement update (given as (17)) contributes more to the PL generation.
- Before $q = 20000$, increasing the reduction order reduces the overestimation effect of zonotopes significantly. After $q = 20000$, the benefit of further increasing the reduction order is not obvious. Moreover, it increases the risk of violating the 10 ms real-time threshold.

Summarizing the results, we can conclude that applying high-dimensional zonotopes requires larger reduction orders, which is more computationally consuming. Therefore, the choice of the zonotope dimension should be prudent and consider the application requirements to avoid wasting computing power.

The performance of the PL estimation is commonly evaluated by applying the Stanford-ESA Integrity Diagram, presented in Fig. 8. The core issue of the PL calculation is to bound the state estimation error, requiring the estimated PL of a state to be larger than the state error. Namely, the sample points shall locate in the top left area in Fig. 8. Further, the estimated PL shall be under a pre-defined limit, namely the alarm limit (AL). Otherwise, the integrity system shall be declared unavailable. Finally, the red area shall be avoided as the highest priority for integrity systems, namely the hazardous operation (also known as hazardously misleading information (HMI)), where the actual state estimation error is beyond the AL, and the estimated PL is under the AL. To conclude, the sample points shall locate in the nominal operation (NO) area, while the estimated PLs should be kept as small as possible.

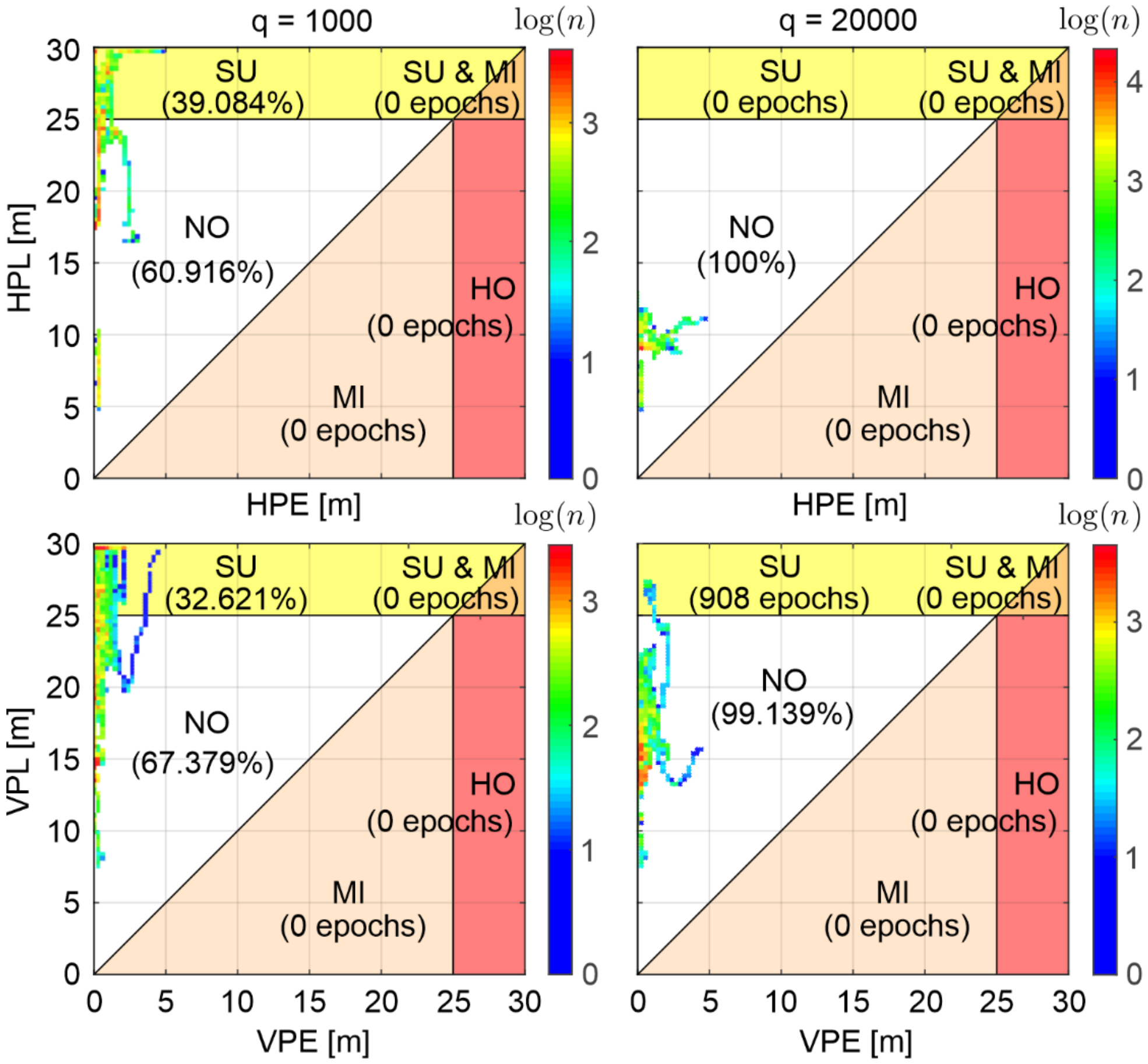

**Fig. 8** Stanford-ESA Integrity Diagram with 105448 epochs. NO = nominal operation; MI = misleading information; HO = hazardous operation; SU = system unavailable. This plot is generated with (Ogier 2022).

The results in Fig. 8 are the position-PLs generated by applying the three-dimensional zonotope. We cannot evaluate the PLs of velocity qualitatively due to insufficient accuracy and precision of the reference, which we demonstrate in the supplementary material. Further, we cannot evaluate the PLs of the orientation due to the missing reference data. Thus, we include the results of the velocity and orientation PLs generated by the nine-dimensional zonotopes in the supplementary material.

Fig. 8 shows that, regardless of the choice of the reduction order $q$, no misleading information is generated by the zonotope, verifying the correctness of the zonotope-based PL estimation. With this result, no integrity risk is observed with the proposed approach. Further, comparing the generated PL by applying $q = 1000$ and $q = 20000$, the overestimation effect with reduction order $q = 1000$ is notable: The sample points are distributed almost evenly along the y-axes, regardless of the positioning error. This also results in system unavailability of up to 39.084%. By applying $q = 20000$, all samples locate in the nominal operation area for HPL estimation.

Table 6 introduces the essential metrics to evaluate the zonotope-based HPL calculation and compares them with the results of the innovation-based method (Zhu et al. 2018), where an EKF applying the same sigma-ε model is developed. The results indicate that the zonotope-based approach outperforms the innovation-based approach in terms of misleading information without compromising on the size of the generated HPL and algorithm availability.

**Table 6** Comparison of the zonotope- and filter-innovation-based PL generation. $P_{mi}$ refers to the probability of misleading information.

| | HPL | | |
|---|---|---|---|
| | Median [m] | $P_{mi}$ [%] | Availability [%] |
| Zonotope-based ($q = 20000$) | **10.71** | **0** | 100 |
| Innovation-based (Zhu et al. 2018) | 10.79 | 0.97 | 100 |

Still, the integrity requirement for safety-critical applications in (EUSPA 2021) is not met with the proposed system, namely HPL under 2 meters. Such a requirement cannot be satisfied by simply increasing the reduction order. Instead, further reducing HPL relies on limiting the measurement uncertainty and, therefore, might be satisfied by integrating further GNSS observations, i.e., carrier phase, into the tightly coupled navigation filter and integrity system.

## Conclusion

This manuscript presented a novel robust state and protection-level estimation scheme within a tightly coupled navigation system involving the EHF and zonotope. The EHF guarantees a robust state estimation under poor parametrization or system disturbances, which was validated with real-world data from mixing environments (including semi/sub-urban environments) in post-processing. By grid searching the same parameter space of the GNSS measurement noise, the 3D positioning accuracy of the EHF varies between 0.6 m and 1 m, whereas the one from the EKF varies between 0.6 m and 2 m. This result proved that the estimation accuracy of the EHF is less sensitive to the GNSS-measurement parametrization and, therefore, has less requirement on the measurement-noise modeling. Applying the same GNSS measurement parametrization, the EHF shows higher robustness against real-world disturbances and outperforms the EKF regarding positioning accuracy, especially in the vertical direction. Considering the difficulty of the appropriate parametrization by applying the EKF within a GNSS/INS, the EHF is suggested to be an ideal alternative to the EKF in urban environments, where the quality of GNSS measurements varies continuously due to changing operating environments. The EHF could benefit UAV applications where vertical localization is essential.

Further, a feasibility study was carried out for the zonotope-based PL generation in a high-order nonlinear, tightly coupled GNSS/INS. In theory, applying zonotope releases the PL generation from the statistical assumptions. The experimental results prove that applying a zonotope for PL generation is valid when the reduction order is appropriately set. Compared to the conventional filter-innovation-based approach, applying zonotope decreases the probability of misleading information to zero and slightly reduces the size of generated PL. This approach is computationally intensive when a large reduction order is required. However, the results prove that a real-time implementation is promising. The size of estimated PLs should be further reduced by applying carrier phase measurements to apply zonotope for safety-critical autonomous applications.

**Acknowledgment**

The authors gratefully acknowledge the contribution of the joint research project Firefly (grant 50NA2103), which is supported by the German Federal Ministry for Economic Affairs and Climate Action (BMWK) in compliance with a resolution of the German Bundestag. Special thanks go to colleagues Jiaying Lin and David Benz for proofreading the article.

## Data Availability

The datasets used and generated in this research are available from the corresponding author upon reasonable request.

## Appendix: H∞ Filtering and Zonotope Theory

For the brevity of the main text, we present the basic theory of the EKF and zonotope in this appendix.

### Theory of Extended H∞ Filter

Considering a nonlinear time-discrete system given as follows:

$$\boldsymbol{x}_k = \boldsymbol{f}(\boldsymbol{x}_{k-1}) + \boldsymbol{G}_{k-1}\boldsymbol{w}_{k-1}, \quad \boldsymbol{w}_k \sim \mathcal{N}(0, \boldsymbol{Q}_k) \tag{A1}$$

$$\boldsymbol{z}_k = \boldsymbol{h}(\boldsymbol{x}_k) + \boldsymbol{v}_k, \qquad \boldsymbol{v}_k \sim \mathcal{N}(0, \boldsymbol{R}_k) \tag{A2}$$

$$\boldsymbol{y}_k = \boldsymbol{L}_k\boldsymbol{x}_k \tag{A3}$$

where $\boldsymbol{x}_k$ and $\boldsymbol{x}_{k-1}$ are the state vectors for time instance $k$ and $k-1$, respectively. $\boldsymbol{z}_k$ is the measurement vector. $\boldsymbol{f}$ and $\boldsymbol{h}$ are the nonlinear time-invariant state transition and measurement functions, respectively. $\boldsymbol{w}_k$ and $\boldsymbol{v}_k$ are the process and measurement noise vectors, respectively. $\boldsymbol{G}_k$ denotes the shaping matrix, which maps the process noise $\boldsymbol{w}_k$ into the state vector $\boldsymbol{x}_k$. $\boldsymbol{w}_k$ and $\boldsymbol{v}_k$ are assumed to satisfy a white Gaussian distribution, whose variance matrices are $\boldsymbol{Q}_k$ and $\boldsymbol{R}_k$, respectively. $\boldsymbol{y}_k$ denotes the desired estimation vector at the time instance $k$, which is defined by a user-designed matrix $\boldsymbol{L}_k$ (full rank). Concerning the navigation solution, all states in $\boldsymbol{x}_k$ shall be estimated such that $\boldsymbol{L}_k$ is defined as a unit matrix. The complete H∞ filtering computation steps are rigorously derived in (Simon 2006) and summarized in (Loo et al. 2019) as:

$$\hat{\boldsymbol{x}}_k^- = \boldsymbol{f}(\hat{\boldsymbol{x}}_{k-1}^+) \tag{A4}$$

$$\hat{\boldsymbol{P}}_k^- = \boldsymbol{F}_{k-1}\hat{\boldsymbol{P}}_{k-1}^+\boldsymbol{F}_{k-1}^T + \boldsymbol{G}_{k-1}\boldsymbol{Q}_{k-1}\boldsymbol{G}_{k-1}^T \tag{A5}$$

$$\hat{\boldsymbol{P}}_k^+ = \hat{\boldsymbol{P}}_k^-\left[\boldsymbol{I} - \gamma\boldsymbol{L}_k^T\boldsymbol{S}_k\boldsymbol{L}_k\hat{\boldsymbol{P}}_k^- + \boldsymbol{H}_k^T\boldsymbol{R}_k^{-1}\boldsymbol{H}_k\hat{\boldsymbol{P}}_k^-\right] \tag{A6}$$

$$\boldsymbol{K}_k = \widehat{\boldsymbol{P}}_k^+ \boldsymbol{H}_k^T \boldsymbol{R}_k^{-1} \tag{A7}$$

$$\widehat{\boldsymbol{x}}_k^+ = \widehat{\boldsymbol{x}}_k^- + \boldsymbol{K}_k\big(\boldsymbol{z}_k - \boldsymbol{h}(\widehat{\boldsymbol{x}}_k^-)\big) \tag{A8}$$

where $\widehat{\boldsymbol{x}}_k^-$ , $\widehat{\boldsymbol{x}}_{k-1}^+$ , $\widehat{\boldsymbol{P}}_k^-$ , and $\widehat{\boldsymbol{P}}_k^+$ denote the a-priori and a-posteriori state and covariance estimate, respectively. $\boldsymbol{F}_{k-1}$ and $\boldsymbol{H}_k$ are the Jacobian matrix of $\boldsymbol{f}$ and $\boldsymbol{h}$, linearized at the operating points, respectively. $\boldsymbol{S}_k$ is a user-defined weighting matrix, specified as a unit matrix in current work, which weights the elements of $\boldsymbol{y}_k$ in the cost function given in (A9). $\boldsymbol{K}_k$ is the filter gain matrix. It should be noted that the calculation of $\boldsymbol{K}_k$ involves a parameter $\gamma$, discussed later in this section.

The purpose of the optimal H∞ filtering is to choose the suitable estimation strategy that minimizes the following cost function:

$$J = \frac{\sum_{k=0}^{N-1} \|\boldsymbol{y}_k - \widehat{\boldsymbol{y}}_k\|_{\boldsymbol{S}_k}^2}{\|\boldsymbol{x}_0 - \widehat{\boldsymbol{x}}_0\|_{\boldsymbol{P}_0^{-1}}^2 + \sum_{k=0}^{N-1}\left(\|\boldsymbol{w}_k\|_{\boldsymbol{Q}_k^{-1}}^2 + \|\boldsymbol{v}_k\|_{\boldsymbol{R}_k^{-1}}^2\right)} \tag{A9}$$

where $\boldsymbol{x}_0$ , $\widehat{\boldsymbol{x}}_0$ and $\boldsymbol{P}_0$ represent the initial state vector, its estimation, and the covariance of the initialization error, respectively. $\widehat{\boldsymbol{y}}_k$ is the estimate of $\boldsymbol{y}_k$. $N$ denotes the current epoch. The cost function $J$ can be understood as the normalization of all the weighted historical estimation errors of $\boldsymbol{y}_k$ from epoch 0 to the last epoch $N - 1$, relative to the initialization error $\boldsymbol{x}_0 - \widehat{\boldsymbol{x}}_0$, the process noise $\boldsymbol{w}_k$ and measurement noise $\boldsymbol{v}_k$ and considering their corresponding covariance $\boldsymbol{P}_0$, $\boldsymbol{Q}_k$ and $\boldsymbol{R}_k$. The choice of the cost function $J$ is based on the game theory approach and is assumed to improve the filter robustness against worst-case situations (Simon 2006, p. 343).

However, the cost function cannot be directly minimized (Simon 2006, p. 344). Thus, a performance bound $\gamma^{-1}$ is introduced as a constraint for the optimization:

$$J < \gamma^{-1} \tag{A10}$$

Solving the constrained optimization problem results in the estimation strategy (A6)~(A8). Meanwhile, the constrained optimization problem is only solvable when the following inequality condition is satisfied (Simon 2006, pp. 351-352):

$$\left(\widehat{\boldsymbol{P}}_k^-\right)^{-1} + \boldsymbol{H}_k^T \boldsymbol{R}_k^{-1} \boldsymbol{H}_k - \gamma \boldsymbol{L}_k^T \boldsymbol{S}_k \boldsymbol{L}_k \succ 0 \tag{A11}$$

where '$\succ$' denotes that the resulting matrix on the left-hand side is positive-definite. It can be observed that the larger is $\gamma$, the more difficult that (A11) can be satisfied. Thus, $\gamma$ shall not be chosen too large. The choice $\gamma$ in the vicinity of *infinity* might cause the divergence of the EHF (Hassibi et al. 1996). On the other hand, Simon (2006) has described $\gamma$ as a measure of filter robustness, i.e. the larger $\gamma$, the smaller estimation error under worst-case situations, and therefore, the stronger robustness of the filter. Thus, the choice of $\gamma$ is crucial and decides the EHF performance. In practice, $\gamma$ could be chosen conservatively by experience or estimated by solving the linear matrix inequality (LMI) problem described by (A11). We apply an LMI solver from MATLAB® to calculate $\gamma$.

## Theory of Zonotope

Zonotope is a type of centrally symmetric convex polytope. Zhang et al. (2020) have applied zonotope to present the bounding area of the estimated states within a filter. This section introduces the definition and necessary background knowledge of zonotope.

**Definition 1.** An $n$-dimensional zonotope with $m$-order $\mathbb{Z} \subset \mathbb{R}^n$ $(n \leq m)$ is an affine transformation of a hypercube $\mathbb{B}^m = [-1,1]^m$ (Zhang et al. 2020):

$$\mathbb{Z} = \{\boldsymbol{\alpha} \in \mathbb{R}^n \mid \boldsymbol{\alpha} = \boldsymbol{c} + \boldsymbol{\mathcal{H}}\boldsymbol{b}, \forall \boldsymbol{b} \in \mathbb{B}^m\} \tag{A12}$$

where vector $\boldsymbol{c} \in \mathbb{R}^n$ is the center of $\mathbb{Z}$. $\boldsymbol{\mathcal{H}} \in \mathbb{R}^{n\times m}$ is called the generation matrix of zonotope $\mathbb{Z}$, which defines the shape and direction of $\mathbb{Z}$. Each column of $\boldsymbol{\mathcal{H}} = [\boldsymbol{h_1}, \dots, \boldsymbol{h_m}]$ denotes a translation direction and amplitude in Euclidean space. For a simple expression, we use $\mathbb{Z} =< \boldsymbol{c}, \boldsymbol{\mathcal{H}} >$ to denote a zonotope. Fig. A1 presents an example of a two-dimensional and six-order zonotope as the yellow area.

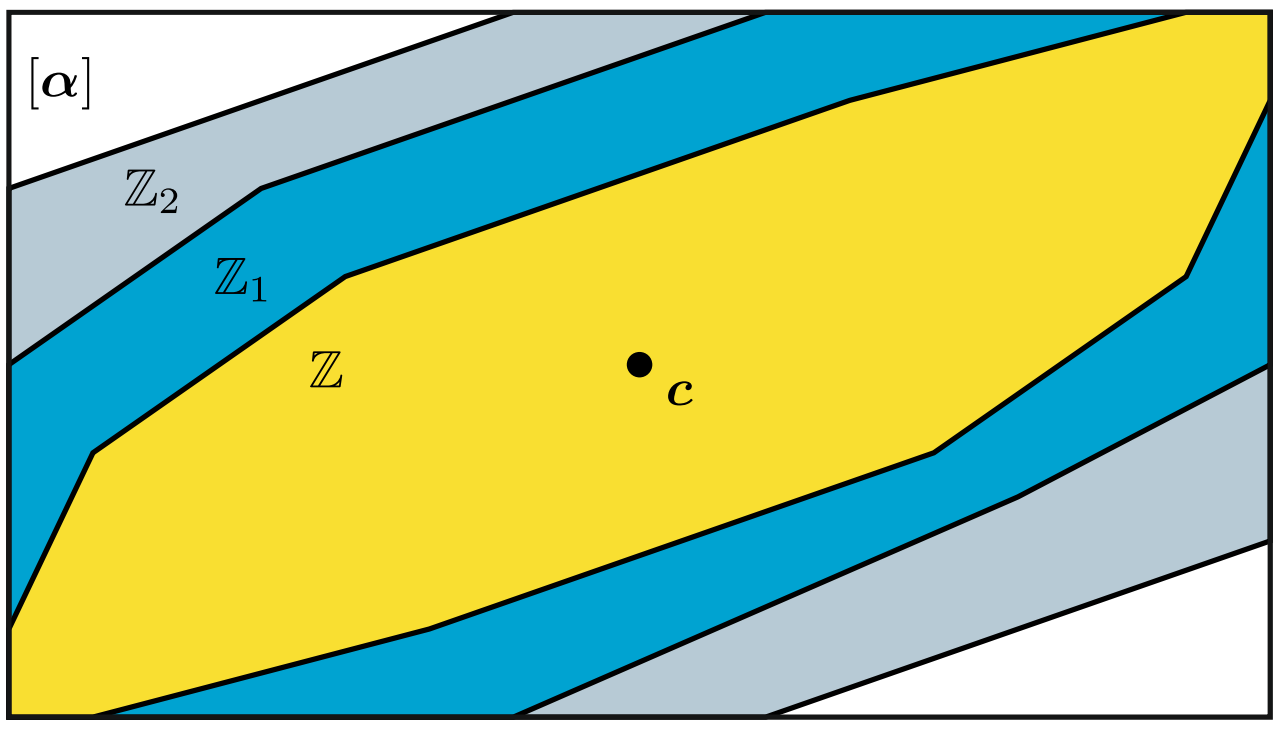

**Fig. A1** An example of a two-dimensional six-order zonotope $\mathbb{Z}$ (yellow). $\mathbb{Z}_1$ (blue) and $\mathbb{Z}_2$ (grey) are reduced with $q_1 = 4$ and $q_2 = 3$, respectively. The outer line is the boundary of the interval hull box $[\boldsymbol{\alpha}]$.

**Property 1.1.** The Minkowski sum of two zonotopes $\mathbb{Z}_1 =< \boldsymbol{c}_1, \boldsymbol{\mathcal{H}}_1 >\subset \mathbb{R}^n$ and $\mathbb{Z}_2 =< \boldsymbol{c}_2, \boldsymbol{\mathcal{H}}_2 >\subset \mathbb{R}^n$ is also a zonotope, which can be computed as:

$$< \boldsymbol{c}_1, \boldsymbol{\mathcal{H}}_1 > \oplus < \boldsymbol{c}_2, \boldsymbol{\mathcal{H}}_2 > = < \boldsymbol{c}_1 + \boldsymbol{c}_2, [\boldsymbol{\mathcal{H}}_1 \ \boldsymbol{\mathcal{H}}_2] > \quad \text{(A13)}$$

**Property 1.2.** The image of a zonotope by a linear transformation $\boldsymbol{\mathcal{L}}$ can be computed by:

$$\boldsymbol{\mathcal{L}} \odot < \boldsymbol{c}, \boldsymbol{\mathcal{H}} > = < \boldsymbol{\mathcal{L}}\boldsymbol{c}, \boldsymbol{\mathcal{L}}\boldsymbol{\mathcal{H}} > \quad \text{(A14)}$$

where $\odot$ denotes the linear image operator.

**Definition 2.** The interval hull box $[\boldsymbol{\alpha}]$ of a zonotope $\mathbb{Z}$ is the smallest centered interval vector containing $\mathbb{Z}$ (Zhang et al. 2020). An example of a two-dimensional zonotope with six-order and its interval hull is depicted in Fig. A1. For an m-order zonotope $\mathbb{Z} =< \boldsymbol{c}, \boldsymbol{\mathcal{H}} >\subset \mathbb{R}^n$ , its interval hull box $[\boldsymbol{\alpha}]$ is computed as:

$$[\alpha_i] = \left[c_i - \sum_{j=1}^{m} \left|\mathcal{H}_{ij}\right|, c_i + \sum_{j=1}^{m} \left|\mathcal{H}_{ij}\right|\right],\ i \in \{1, \dots, n\} \quad \text{(A15)}$$

**Remark 1.** When the outer approximation of a zonotope $\mathbb{Z}$ by an aligned box is too conservative, a reduction operator $\mathcal{R}(\boldsymbol{\mathcal{H}})$ can be used to reduce the complexity of $\mathbb{Z}$ (Combastel 2016). After the reduction, the original zonotope is a subset of the resulting zonotope,

$$< \boldsymbol{c}, \boldsymbol{\mathcal{H}} > \subseteq < \boldsymbol{c}, \mathcal{R}(\boldsymbol{\mathcal{H}}) > \quad \text{(A16)}$$

without changing the interval hull box. The reduction order $q$ ($n \leq q \leq m$) specifies the order of the resulting zonotope $< \boldsymbol{c}, \mathcal{R}(\boldsymbol{\mathcal{H}}) >$, where $\mathcal{R}(\boldsymbol{\mathcal{H}}) \in \mathbb{R}^{n \times q}$. The realization of zonotope reduction is given in (Zhang et al. 2020).

In Fig. A1, an example of reducing the two-dimensional zonotope ($\mathbb{Z} =< \boldsymbol{c}, \boldsymbol{\mathcal{H}} > \subset \mathbb{R}^2$) with six-order ($\boldsymbol{\mathcal{H}} \in \mathbb{R}^{2\times 6}$) is illustrated. $\mathbb{Z}$ is reduced to $\mathbb{Z}_1$ and $\mathbb{Z}_2$ with reduction order $q_1 = 4$ and $q_2 = 3$, respectively. A smaller reduction order $q$ results in a larger overestimation but reduces the computational load. The reduction order $q$ is allowed to be freely chosen (Combastel 2003), compromising between the exactness (with larger $q$) and the domain complexity (with smaller $q$) of the resulting zonotope.

## Author Biography

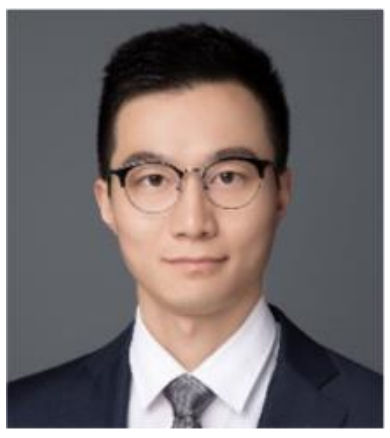

**Shuchen Liu** received his B. Sc. degree from Shanghai Jiao Tong University in 2014 and his M. Sc. degree from RWTH Aachen University in 2017, both in automation engineering. Since 2017, he has been a research associate and pursuing his Ph.D. degree in the Navigation research group at the Institute of Automatic Control. His research focuses on robust state estimation and integrity monitoring of GNSS-based navigation systems for different applications, such as automobiles and vessels.

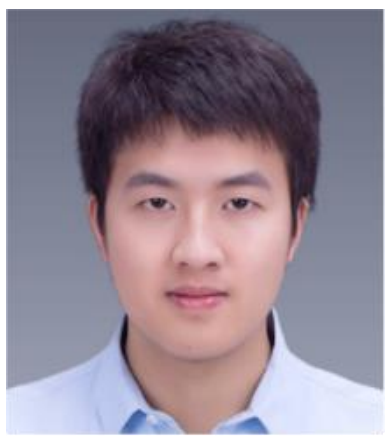

**Kaizheng Wang** received his B.Eng. degree in mechatronics engineering from Zhejiang University in 2017 and M. Sc. degree in automation engineering from RWTH Aachen University in 2020. Currently, he is a Ph.D. candidate in the Department of Computer Science, M-Group, Campus Bruges, KU Leuven. His research activities focus on uncertainty modeling and optimization and epistemic artificial intelligence.

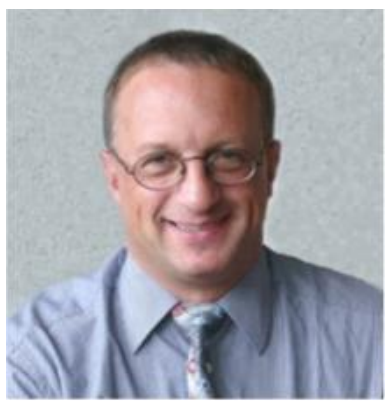

**Dirk Abel** received his Dipl.-Ing degree in mechanical engineering in 1983, and the Dr.-Ing. Degree in 1987 from RWTH Aachen University. There, in 2001, he became a full professor and head of the Institute of Automatic Control. From 2009 to 2017, he was a member of the supervisory board of KUKA AG, Germany. Professor Abel's main research area covers model predictive control, nonlinear and robust control, and rapid control prototyping in different application areas.